# Comparison and Evaluation of Isotope-Ratio Calculation Schemes for Compound-Specific Chlorine Isotope Analysis Using GC-EI-MS

**Caiming Tang[a,b,]* and Lianjun Bao[c]**

[a] *State Key Laboratory of Organic Geochemistry, Guangzhou Institute of Geochemistry, Chinese Academy of Sciences, Guangzhou 510640, China*

[b] *University of Chinese Academy of Sciences, Beijing 100049, China*

[c] *School of Environment, Jinan University, Guangzhou 510632, China*

*Corresponding Author.

Tel: +86-020-85291489; Fax: +86-020-85290009; E-mail: CaimingTang@gig.ac.cn (C. Tang).

A recent article reported a comparison study concerning compound-specific chlorine isotope analysis (CSIA-Cl) of organochlorines using gas chromatography- isotope ratio mass spectrometry (GC-IRMS) and gas chromatography-quadrupole mass spectrometry (GC-qMS).[1] Comparable precisions between the two instruments were achieved and trueness of the analysis results was confirmed. The CSIA-Cl method using GC-qMS was originally developed by Sakaguchi-Söder et al.[2] and further improved by several relevant studies,[3,4,5] and has been used to evaluate organic contaminant attenuation in the environment.[6,7] The essential principle of the method is to calculate the chlorine isotope ratio ($^{37}Cl/^{35}Cl$) by using the first pair of neighboring chlorine isotopologues (two-mass apart) of either the molecular ion or a fragmental ion of a target analyte, or using all the first isotopologue pairs of all molecular and fragmental ions. These isotope-ratio calculation schemes using chlorine isotopologue pair(s) were based on the binomial theorem and the prerequisite hypothesis that the measured abundances of chlorine isotopologues of individual analytes were binomially distributed. However, it is not true with the hypothesis.

Recently, we have experimentally and theoretically confirmed that the chlorine isotopologues of detected ions of individual organochlorines on electron ionization-MS (EI-MS) did not conform to binomial distribution.[8] In addition, the chlorine isotopologues of synthetic organochlorines with more than one Cl atom produced under some conditions and all environmental polychlorinated organic compounds are unlikely exactly binomial distributed.[9] Therefore, the isotopologue-pair schemes of isotope-ratio calculation derived from the binomial distribution of chlorine isotopologues of detected ions of individual analytes are procedurally incorrect, even though the achieved data could reflect the trueness in some cases.[1,4,5] This scenario may be an example of near miss to approach some correct conclusions or results, but it is wrong indeed.

Moreover, in our study, the isotope-ratio variation tendencies of perchlorethylene (PCE) standards from two sources were inconsistent when two different isotope-ratio calculation schemes were applied (Figure 1), which were a complete-isotopologue scheme using all isotopologues of the molecular ion of PCE and an isotopologue-pair scheme using the first pair of isotopologues of the molecular ion. This resulted in obviously contradictory relative variations of isotope ratios obtained with different schemes when a standard from a source was used as the external isotopic standard for the standard from another source. To our point's view, the complete-isotopologue scheme rather than the isotopologue-pair schemes should be better proposed for CSIA-Cl when the detection is performed by GC-EI-MS. When using the isotopologue-pair schemes, analysts should take account of the specific isotopologue distributions of external isotopic standards and analytes, which may trigger biased results in CSIA-Cl even though two-point calibration is conducted.

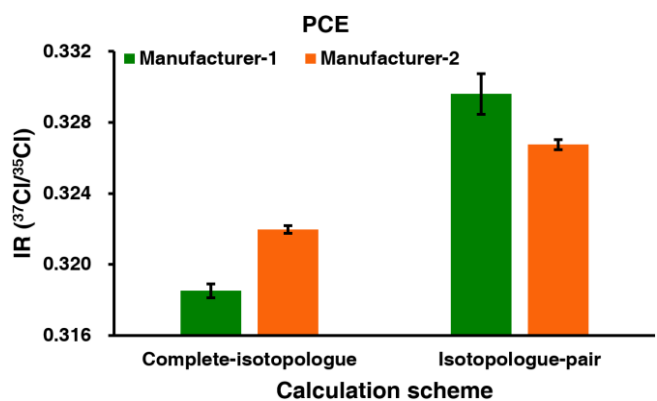

**Figure 1**. Measured isotope ratios of perchlorethylene (PCE) standards from different manufacturers calculated with different isotope-ratio calculation schemes. IR: isotope ratio. Error bars denote the standard deviations (1σ, n=5). The instrument used was GC-double focus magnetic-sector high resolution MS (GC-DFS-HRMS).

The complete-isotopologue scheme and the previously reported complete-ion method can certainly avoid these drawbacks caused by the source-specific isotopologue distributions of organochlorines for CSIA-Cl.[10,11] Technically, the complete-isotopologue scheme and the complete-ion method may not suitable for CSIA-Cl of organochlorines with more than two Cl atoms using GC-qMS, due to its insufficient sensitivity for the analytes.[11] Alternatively, GC-double focus magnetic-sector high resolution MS (GC-DFS-HRMS) can provide sufficient signal intensities for organochlorines with the number of Cl atoms up to at least four using the complete-isotopologue scheme with fairly low injection amounts (around 1 ng on column).[10] Consequently, GC-DFS-HRMS in association with the complete-isotopologue scheme of isotope-ratio calculation can be a promising alternative approach for performing CSIA-Cl.

**ACKNOWLEDGEMENTS**

The authors are grateful for Mr. Jianhua Tan, from Guangzhou Quality Supervision and Testing Institute, China for his help in instrumental analysis. This study was partially financed by the National Natural Science Foundation of China (Grant No. 41603092).

**Notes**

The authors declare no competing financial interest.